\documentclass[aps,pr,superscriptaddress,numerical,preprint,showkeys,floatfix]{revtex4-1}

\usepackage{epsfig,amsmath,amssymb,txfonts,hyperref}

\begin{document}

\title{Geometries of edge and mixed dislocations in bcc Fe from first principles calculations}

\author{Michael R. Fellinger}
\email{mfelling@illinois.edu}
\affiliation{Department of Materials Science and Engineering, University of Illinois at Urbana-Champaign, Urbana, Illinois 61801, USA}
\author{Anne Marie Z. Tan}
\affiliation{Department of Materials Science and Engineering, University of Illinois at Urbana-Champaign, Urbana, Illinois 61801, USA}
\affiliation{Department of Materials Science and Engineering, University of Florida, Gainesville, Florida 32611, USA}
\author{Louis G. Hector Jr.}
\affiliation{General Motors Global R\&D Center, 30500 Mound Road, Warren, MI 48092, USA}
\author{Dallas R. Trinkle}
\affiliation{Department of Materials Science and Engineering, University of Illinois at Urbana-Champaign, Urbana, Illinois 61801, USA}

\date{\today}
\begin{abstract}
We use density functional theory (DFT) to compute the core structures of $a_0[100](010)$ edge, $a_0[100](011)$ edge, 
$a_0/2[\bar{1}\bar{1}1](1\bar{1}0)$ edge, and $a_0/2[111](1\bar{1}0)$ $71^{\circ}$ mixed dislocations in body-centered cubic (bcc) Fe. The 
calculations are performed using flexible boundary conditions (FBC), which effectively allow the dislocations to relax as isolated defects by 
coupling the DFT core to an infinite harmonic lattice through the lattice Green function (LGF). We use the LGFs of the dislocated geometries in 
contrast to most previous FBC-based dislocation calculations that use the LGF of the bulk crystal. The dislocation LGFs account for changes in the 
topology of the crystal in the core as well as local strain throughout the crystal lattice. A simple bulk-like approximation for the force constants 
in a dislocated geometry leads to dislocation LGFs that optimize the core structures of the $a_0[100](010)$ edge, $a_0[100](011)$ edge, and 
$a_0/2[111](1\bar{1}0)$ $71^{\circ}$ mixed dislocations. This approximation fails for the $a_0/2[\bar{1}\bar{1}1](1\bar{1}0)$ dislocation however, so 
in this case we derive the LGF from more accurate force constants computed using a Gaussian approximation potential. The standard deviations of the 
dislocation Nye tensor distributions quantify the widths of the dislocation cores. The relaxed cores are compact, and the local magnetic moments on 
the Fe atoms closely follow the volumetric strain distributions in the cores. We also compute the core structures of these dislocations using eight 
different classical interatomic potentials, and quantify symmetry differences between the cores using the Fourier coefficients of their Nye tensor 
distributions. Most of the core structures computed using the classical potentials agree well with the DFT results. The DFT core geometries provide 
benchmarking for classical potential studies of work-hardening, as well as substitutional and interstitial sites for computing solute-dislocation 
interactions that serve as inputs for mesoscale models of solute strengthening and solute diffusion near dislocations.
\end{abstract}


\maketitle

\section{Introduction}

Steel alloys are used in a wide variety of structural applications due to their low cost and the relative ease of tuning their mechanical properties 
via alloying and processing compared to many other structural materials~\cite{leslie91,berns08}. The ferrite phase found in many steels is 
body-centered cubic (bcc) Fe containing C and other solute atoms~\cite{leslie91,berns08,devaraj18}. As in other bcc metals, dislocation slip is one of 
the most important plastic deformation mechanisms in bcc Fe~\cite{christian83, taylor92}. Therefore, accurate modeling of dislocation structures in Fe 
and their response to stress is key to understanding deformation behavior, improving microstructure-based models of plasticity and fracture, and 
ultimately developing new steels with improved mechanical properties. The $a_0/2 \langle 111 \rangle$-type screw dislocations in bcc metals have been 
widely studied since these dislocations largely control the low-temperature plastic deformation of bcc metals and alloys~\cite{christian83, duesbery89, 
taylor92}. The details of the screw dislocation core structure are known to affect the Peierls stress and therefore the mobility of these 
dislocations~\cite{Cai2004, Chaussidon2006, gordon2010}, and density functional theory (DFT) calculations first revealed that the core is compact and 
symmetric compared to the degenerate core structure predicted by many classical interatomic 
potentials~\cite{ismail2000,woodward2001,woodward2002,frederiksen03}.  The questionable reliability of classical potentials and the lack of experimental 
measurements of dislocation core structures in Fe highlights the need for electronic structure methods to compute detailed atomic-level 
structural features in dislocation cores.

While $a_0/2 \langle 111 \rangle$ screw dislocations predominantly govern the plastic response of bcc metals at low temperatures, 
dislocations of edge or mixed character may also play important roles in controlling plastic deformation in bcc metals. For example, edge 
dislocations in bcc metals can form from reactions of dislocations with $a_0/2 \langle 111 \rangle$-type Burgers vectors. As screw dislocations move through the material, they can react with other dislocations intersecting their glide plane and form stable binary junctions with Burgers vector $a_0 \langle 100 \rangle$ via a reaction of type \cite{puschl1985,schoeck2010}
\begin{equation}
a_0/2 [111] + a_0/2 [1\bar{1}\bar{1}] \rightarrow a_0 [100].
\end{equation}
These binary junctions may themselves be mobile, or further react with other dislocations to form ternary junctions which contribute to work 
hardening. These junction reactions are of interest and have been studied by dislocation dynamics simulations~\cite{bulatov2006,madec2008}. 
Here, we consider two possible edge dislocations with $a_0 \langle 100 \rangle$-type Burgers vectors---$a_0 \langle 100 \rangle \{010\}$ and 
$a_0 \langle 100 \rangle \{011\}$---along with a $a_0/2 \langle 111 \rangle \{011\}$ edge dislocation, as this is the
most commonly observed type of edge dislocation in bcc Fe~\cite{clouet08}. Edge dislocations are also of interest for understanding the 
influence of dislocation loops~\cite{bonny16, fikar17} and cell structures~\cite{haghighat14a} on deformation processes in Fe. 
Experimentally observed edge dislocations in nanocrystalline samples of bcc W~\cite{wei2006} and Ta~\cite{wei2011} are believed to be the primary 
reason for the reported lower strain rate sensitivity of nanocrystalline bcc metals and alloys compared to their coarse-grained counterparts, and
may play an important role in controlling the plastic response of nanocrystalline bcc Fe.
Finally, dislocations in bcc Fe can play a part in other interesting phenomena as well. For example, pipe diffusion (i.e. accelerated diffusion along 
the dislocation line) of C interstitials has been predicted to occur in the $a_0/2[111](1\bar{1}0)$ $71^{\circ}$ mixed dislocation in bcc Fe 
\cite{ishii2013}. However, straightforward pipe diffusion was not predicted for other types of dislocations---the migration of C interstitials was 
found to be accelerated not along the dislocation line direction but in a conjugate diffusion direction formed by a pathway of octahedral interstitial
sites adjacent to the dislocation core. In order to better understand the complex mechanisms that are likely to be at play here, accurate and detailed 
descriptions of the dislocation cores are necessary.

In this study, we use DFT combined with flexible boundary conditions (FBC)~\cite{sinclair78,rao98,woodward05} to optimize the core structures of the 
$a_0[100](010)$ edge, $a_0[100](011)$ edge, $a_0/2[\bar{1}\bar{1}1](1\bar{1}0)$ edge, and $a_0/2[111](1\bar{1}0)$ $71^{\circ}$ mixed dislocations in 
bcc Fe. Previous simulations of edge and mixed dislocations in bcc Fe have relied on classical interatomic potentials due to the large supercells 
needed to contain the long-ranged strain fields generated by dislocations~\cite{ishii2013, chen00, clouet08, monnet09, queyreau11, terentyev11, 
wang13, swinburne13, bhatia14, haghighat14a, haghighat14b, bonny16, anento18}. Yan et al.~\cite{yan04} and Chen et al.~\cite{chen06} used 
first-principles calculations to study the electronic effects of C solutes and kinks on edge dislocations in bcc Fe, respectively. However, both of 
these studies used a Finnis-Sinclair classical potential to generate the initial dislocation geometries for the first-principles calculations. The 
accuracy of results from classical simulations strongly depends on the fidelity of the interatomic potential, and there are no experimental 
measurements or first-principles calculations of edge and mixed dislocation core structures in bcc Fe to benchmark the core structures from classical 
potentials. We therefore present the first fully {\it{ab initio}} calculations of the core structures of edge and mixed dislocations in bcc Fe. Our 
DFT-based FBC calculations allow a single dislocation to effectively relax as an isolated defect in a supercell size tractable for DFT calculations 
by coupling the DFT core to an infinite harmonic lattice through the lattice Green function (LGF)~\cite{sinclair78,rao98,woodward05,trinkle08}. In 
contrast to most previous DFT-based FBC calculations of dislocation cores that used the LGF of the bulk crystal to approximate the LGF of dislocated 
geometries, here we use LGFs specifically computed for each dislocation~\cite{tan16}. The dislocation LGFs account for changes in both the topology 
of the crystal lattice in the highly-distorted core region and local strain throughout the lattice. The FBC method removes any reliance on 
dislocation multipole arrangements often used in DFT simulations to cancel the long-ranged strain fields generated by dislocations, but that may 
generate artifacts in the dislocation core structures due to dislocation-dislocation interactions. Our DFT core structures serve to benchmark the 
predictions of existing potentials, provide fitting data for generating new classical potentials, serve as a basis of comparison for future 
experimental investigations of dislocation cores in bcc Fe, and also serve as the starting point for first-principles calculations of solution 
strengthening~\cite{yasi10} and solute transport near edge and mixed dislocations in bcc Fe~\cite{schiavone16}.

The rest of this paper is organized as follows. Section~\ref{sec:Methods} presents our computational geometries, discusses the FBC method, and gives 
the details of our DFT calculations. Here we discuss how we visualize the dislocation cores using a combination of differential displacement 
maps~\cite{vitek70}, Nye tensor distributions~\cite{hartley05a, hartley05b}, volumetric strain, and changes in the magnetic moments on the Fe atoms. 
This section also presents how we quantify the widths of the dislocation cores using the second moments of the Nye tensor distributions, and how we 
distinguish symmetry differences between the core structures from DFT and classical potentials using the Fourier coefficients of the Nye tensor 
distributions. Section~\ref{sec:Results} presents our DFT-optimized dislocation cores, and compares the results to core structures optimized using 
eight different interatomic potentials. Section~\ref{sec:Summary} summarizes our results and provides further discussion.

\section{Computational methods}\label{sec:Methods}

\subsection{First principles calculations with flexible boundary conditions}\label{subsec:DFT-FBC}

\begin{figure}[htb]
\centering
\includegraphics[width=3.00in]{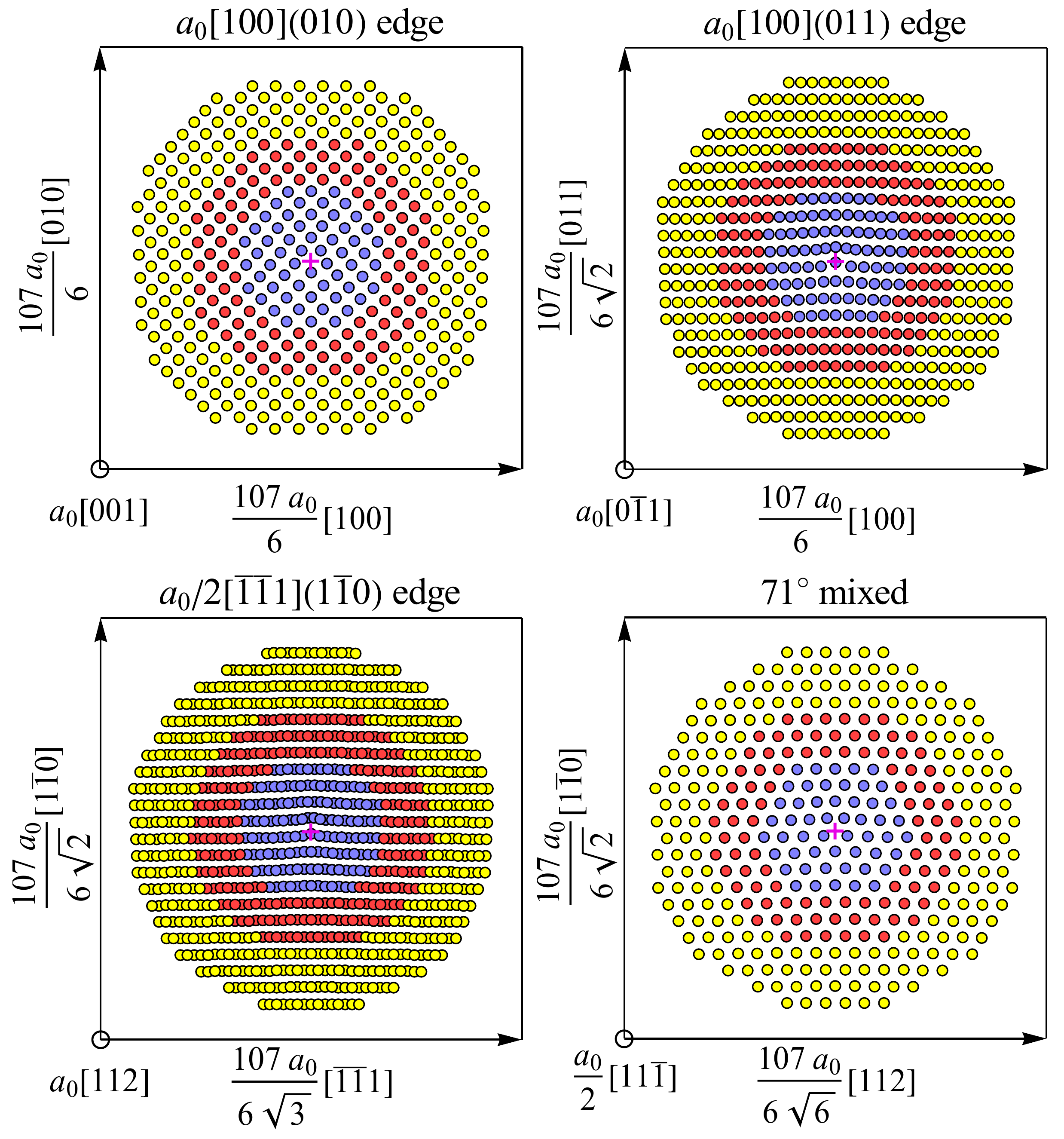}
\caption{(color online). Initial supercell geometries for the $a_0[100](010)$ edge, $a_0[100](011)$ edge, $a_0/2[\bar{1}\bar{1}1](1\bar{1}0)$ edge, 
and $a_0/2[111](1\bar{1}0)$ $71^{\circ}$ mixed dislocations in bcc Fe. The lattice parameter $a_0 = 2.832$ {\AA}, and the supercell dimensions 
perpendicular to the dislocation threading direction are 50.46 {\AA}. The atoms are displaced according to anisotropic elasticity theory and divided 
into three regions to apply FBC. The magenta ``+" marks the center of the elastic displacement field. The atoms are 
surrounded by a vacuum region in all four cases since the dislocation displacement fields are incompatible with periodic boundary conditions. Each 
supercell is subject to periodic boundary conditions along the threading direction. Table~\ref{tab:geometry} provides more details about the 
dislocation geometries.}
\label{fig:initial-geometries}
\end{figure}

Figure~\ref{fig:initial-geometries} shows the initial dislocation geometries that we optimize using first-principles calculations with 
FBC. We construct cylindrical slab geometries and introduce the dislocations by displacing all the atoms in the slabs according to the 
displacement fields predicted by anisotropic elasticity theory~\cite{bacon80}. The magenta ``+" symbols in the figure show the center of the elastic 
displacement field for each dislocation. The displacement fields of edge and mixed dislocations are incompatible with periodic boundary conditions 
perpendicular to the dislocation threading direction (pointing out of the page), so we surround each slab by a vacuum region. We divide each slab 
into region 1 (blue), region 2 (red), and region 3 (yellow) for applying FBC which we discuss in the next paragraph. The supercell dimensions 
perpendicular to the threading directions are equal for all the dislocations, with dimensions of 50.46 {\AA} $\times$ 50.46 {\AA}. Each supercell is 
periodic along the threading direction which requires that the slabs have different thicknesses along this direction. The radial thickness of region 
2 is determined by the interaction range of atoms in bcc Fe, and the radial thickness of region 3 is chosen large enough to isolate regions 1 and 2 
from the effects of the vacuum. We chose the radial thickness of region 1 large enough to ensure that the highly-distorted dislocation cores are 
confined to region 1, which is confirmed by the differential displacement maps and Nye tensor distributions in 
Figs.~\ref{fig:edge100}-\ref{fig:mixed}. Table~\ref{tab:geometry} gives the radii and numbers of atoms for each region.

\begin{table}
\caption{\label{tab:geometry} Geometry information for the $a_0[100](010)$ edge, $a_0[100](011)$ edge, $a_0/2[\bar{1}\bar{1}1](1\bar{1}0)$ edge, and 
$a_0/2[111](1\bar{1}0)$ $71^{\circ}$ mixed dislocations in bcc Fe. The table lists the number of atoms and the radius in {\AA} of region 1 (blue atoms 
in Fig.~\ref{fig:initial-geometries}), region 2 (red atoms in Fig.~\ref{fig:initial-geometries}), and region 3 (yellow atoms in 
Fig.~\ref{fig:initial-geometries}). The radius of each region is nearly equal for each dislocation, but the number of atoms in each region varies 
between the dislocations due to the different slab thicknesses along the dislocation threading direction.\newline}
\centering
{\renewcommand{\arraystretch}{1.0}
\begin{tabular}{ccccccc} \hline\hline
\multicolumn{1}{c}{} &
\multicolumn{2}{c}{region 1} &
\multicolumn{2}{c}{region 2} &
\multicolumn{2}{c}{region 3} \\
dislocation                                & atoms & radius & atoms & radius & atoms & radius \\ \hline
$a_0[100](010)$ edge                       & 60    & 8.8    & 110   & 14.7   & 216   & 22.4 \\
$a_0[100](011)$ edge                       & 82    & 8.7    & 150   & 14.6   & 300   & 22.9 \\
$a_0/2[\bar{1}\bar{1}1](1\bar{1}0)$ edge   & 142   & 8.7    & 261   & 14.5   & 514   & 21.8 \\
$a_0/2[111](1\bar{1}0)$ $71^{\circ}$ mixed & 52    & 8.8    & 96    & 14.8   & 190   & 22.4 \\
\hline\hline
\end{tabular}}
\end{table}

The FBC approach~\cite{sinclair78, woodward05} couples the highly distorted dislocation core to an infinite harmonic bulk which 
effectively allows a dislocation to relax as an isolated defect. The FBC approach consists of two steps: in the first 
step we use a conjugate gradient optimization scheme with DFT-computed forces to relax the defect core (region 1), while holding the rest of the 
atoms fixed. This reduces the forces in region 1 but induces forces in region 2. In the second step, we apply displacements on all atoms in regions 
1, 2 and 3 in response to the forces in region 2, as prescribed by the LGF $G$,
\begin{equation}\label{eqn:LGFdisplacements}
\mathbf{u}(\mathbf{R}') = \sum_{\mathbf{R}} G(\mathbf{R} - \mathbf{R}') \mathbf{f}(\mathbf{R}),
\end{equation}
where $\mathbf{u}(\mathbf{R}')$ is the displacement vector of the atom at $\mathbf{R}'$ and $\mathbf{f}$ is the Hellmann-Feynman force on the 
atom at $\mathbf{R}$. The LGF is the pseudoinverse of the force constant matrix $D$~\cite{trinkle08},
\begin{equation}
\sum_{\mathbf{R}''} D(\mathbf{R} - \mathbf{R}'') G(\mathbf{R}'' - \mathbf{R}') = \mathbf{1}\delta(\mathbf{R} - \mathbf{R}'),
\end{equation}
where the force constant matrix element $D_{ab}$ between the atoms at $\mathbf{R}$ and $\mathbf{R}'$ is
\begin{equation}
D_{ab}(\mathbf{R} - \mathbf{R}') = \left. \frac{\partial^2 U^{\mathrm{total}}}{\partial u_a (\mathbf{R}) \partial u_b (\mathbf{R}')} 
\right|_{\mathbf{u}=0}.
\end{equation}
Here $U^{\mathrm{total}}$ is the total potential energy of the crystal, and $u_a$ and $u_b$ are Cartesian components of the displacement vectors. The 
displacements given by the LGF describe the response of an infinite harmonic system; since our system deviates from this harmonic approximation 
particularly in the dislocation core, this generates forces in region 1. Therefore, we alternate between these two steps until all forces in regions 
1 and 2 are smaller than a defined tolerance. We use an efficient numerical method developed in Ref.~\cite{tan16} to compute the LGFs from the force 
constants in the dislocated geometries. The force constant and LGF calculations are discussed in the following paragraphs, and the details of the DFT 
calculations are discussed in Section~\ref{sec:DFT}.

We compute the force constants for the $a_0[100](010)$ edge, $a_0[100](011)$ edge, and $a_0/2[111](1\bar{1}0)$ $71^{\circ}$ mixed dislocations using 
the bulk-like approximation described in Ref.~\cite{tan16}. We use the small displacement method~\cite{kresse95, alfe01, alfe09} to compute the force 
constants of perfect bulk bcc Fe (see Sec.~\ref{sec:DFT} for details). We then approximate the force constants between pairs of atoms in the 
dislocation geometries by assigning to them the force constants from the pair of atoms in the bulk which have the closest equivalent pair vector. We 
have found that this simple approximation works well for most dislocations in simple crystal structures since the force constants are short-ranged 
and the local environment of atoms appears bulk-like even close to the core~\cite{tan16}.

\begin{figure}[htb]
\centering
\includegraphics[width=3.00in]{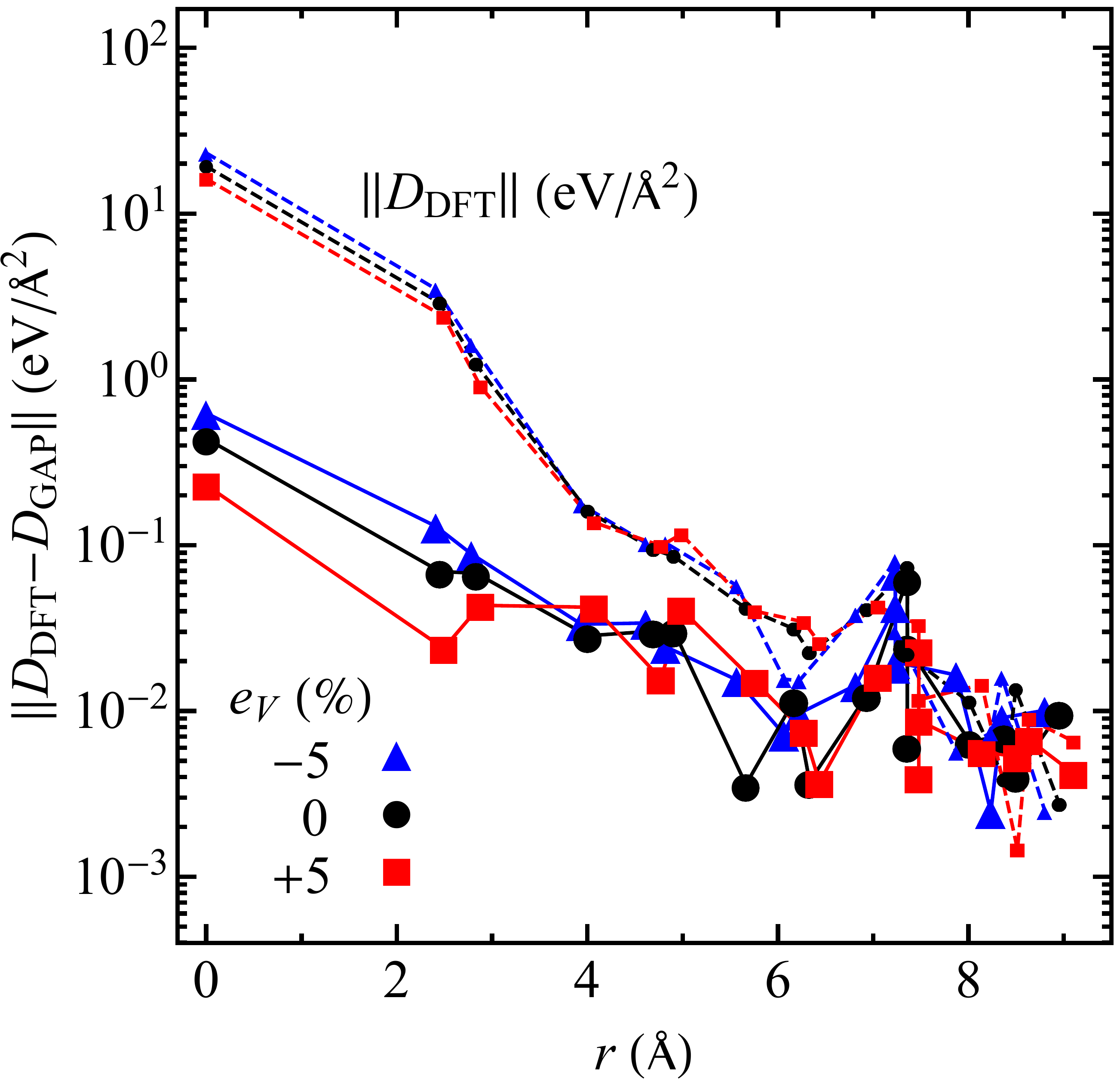}
\caption{(color online). Difference between the DFT and GAP force constants versus distance $r$ between pairs of atoms for different volumetric 
strains $e_V$. For each value of $r$ we take the Frobenius norm of the difference between the DFT and GAP force constant matrices $D$. The dashed 
lines show $||D_{\mathrm{DFT}}||$ for each value of strain. The force constants decay as $r$ increases and the differences show similar behavior. The 
maximum difference for each strain occurs at $r = 0$ (i.e., the on-site term), where $||D_{\mathrm{DFT}}|| = 19.273$ eV/{\AA}$^2$ and 
$||D_{\mathrm{GAP}}|| = 19.713$ eV/{\AA}$^2$ for $e_V = 0$. The corresponding relative error is 2.28\%, with similar maximum errors for $e_V = -5$\% 
(2.69\% error) and $e_V = +5$\% (1.42\% error).}
\label{fig:FC-vs-vol-V9}
\end{figure}

We use a Gaussian approximation potential (GAP) for bcc Fe~\cite{dragoni18} to compute the force constants for the 
$a_0/2[\bar{1}\bar{1}1](1\bar{1}0)$ edge dislocation. For this dislocation, the bulk-like approximation failed to produce adequate force constants. 
This appears to be due to atoms in the initial dislocation core geometry being too close, making it difficult to correctly determine the appropriate 
pairs of atoms in the bulk corresponding to pairs of atoms in the dislocation. Therefore, we compute the force constant matrix for this dislocation 
using a finite-difference scheme on each atom in the dislocation geometry to compute derivatives in forces. Since it is prohibitively expensive to do 
so with DFT, we instead use the Fe GAP to compute the dislocation force constants. The GAP method~\cite{bartok10} generates classical interatomic 
potentials that accurately interpolate the potential energy surface of a material using highly flexible basis functions called ``smooth overlap of 
atomic positions'' (SOAP) kernels. The SOAP kernels can represent a large range of different local atomic environements that can be encountered 
during atomistic simulations, and the accuracy and transferability of GAP steps from fitting the SOAP coefficients to a large set of DFT energies, 
forces, and virials that capture the potential energy surface. We chose the GAP potential for computing the large number of force constants in the 
$a_0/2[\bar{1}\bar{1}1](1\bar{1}0)$ edge dislocation geometry since it provides a good balance between accuracy and speed---while orders of magnitude 
slower than EAM or MEAM, GAP is still much faster than DFT and can provide accuracy comparable to DFT for computing the properties of bcc 
Fe~\cite{dragoni18}. We check that the GAP accurately reproduces the lattice and elastic properties from DFT, which is important to ensure 
consistency between the DFT and LGF relaxations. The GAP lattice constant for bcc Fe is $a_0$ = 2.834 {\AA} and the elastic constants are $C_{11} = 
285.9$ GPa, $C_{12} = 154.3$ GPa and $C_{44} = 103.8$ GPa, which agree well with our DFT-computed lattice constant of $a_0$ = 2.832 {\AA} and elastic 
constants $C_{11} = 277.5$ GPa, $C_{12} = 147.7$ GPa and $C_{44} = 98.1$ GPa~\cite{Fellinger2017}. In addition, we check that the force constants 
from the GAP agree well with the force constants from DFT. Figure \ref{fig:FC-vs-vol-V9} compares the DFT and GAP force constants computed for bulk 
bcc Fe under different volumetric strains $e_V$. The maximum absolute errors between the GAP and DFT force constants occur for the on-site term 
$(r=0)$, which correspond to relative errors of less than 3\% for all three strain values, $e_V = -5$\%, $e_V = 0$\% and $e_V = +5$\%. Therefore, we 
expect the GAP to predict force constants in the strained dislocation geometries which are consistent with DFT.

We numerically invert the dislocation force constant matrices following the method developed in Ref.~\cite{tan16}. This method requires setting up a 
large system divided into five regions: regions 1, 2, and 3 which make up the DFT supercell, a buffer region, and a far-field region. The far-field region 
contains atoms far away from the core whose displacements we approximate using the bulk elastic Green function (EGF) which is the known large 
distance limit of the LGF \cite{trinkle08,yasi12}, while the buffer region contains the remaining atoms between region 3 and the far-field. For all 
the dislocations studied here, we used a buffer size of at least $20 a_0$, for which the errors in the LGF computation due to the far-field 
approximation are on the order of $10^{-3}\textrm{\AA}^2$/eV or less. We compute the LGF for forces in region 2 by applying a unit force on an atom 
in region 2, evaluating the resulting far-field displacements based on the EGF, determining the forces these displacements generate in the buffer 
region, and finally solving for the displacement field corresponding to the effective forces in the system by using a conjugate gradient method to 
numerically invert the force constant matrix. This gives one column of the LGF; by systematically looping through every atom in region 2, we compute 
the LGF matrix that gives displacements on atoms in regions 1, 2, and 3 due to forces in region 2. For more details on this method, the reader is 
referred to Ref.~\cite{tan16}.

\subsection{Density functional theory calculation details}\label{sec:DFT}

We use the plane-wave basis DFT code {\textsc {vasp}}~\cite{kresse96} to generate data for computing bulk force constants and to optimize the 
geometries of the edge and mixed dislocations in bcc Fe. The Perdew-Burke-Ernzerhof (PBE) generalized gradient approximation (GGA) 
functional~\cite{perdew96} accounts for electron exchange and correlation energy, and a projector augmented wave (PAW) potential~\cite{blochl94a} 
with electronic configuration [Ar]$3 d^7 4 s^1$ generated by Kresse and Joubert~\cite{kresse99} models the Fe nuclei and core electrons. The 
calculations require a plane-wave energy cutoff of 400 eV to converge the energies to less than 1 meV/atom. We ensure accurate forces for force 
constant calculations and atomic relaxation using Methfessel-Paxton smearing\cite{methfessel89} with an energy smearing width of 0.25 eV. We chose 
this smearing width to ensure close agreement between the smeared electronic density of states (DOS) of bulk bcc Fe near the Fermi energy and the DOS 
computed using the linear tetrahedron method with Bl{\"o}chl corrections~\cite{blochl94b}. The energy tolerance for the electronic self-consistency 
cycle is $10^{-8}$ eV. All of the calculations are spin polarized to model the ferromagnetism of bcc Fe.

We use the small displacement method~\cite{kresse95, alfe01, alfe09} to compute the force constants of bulk bcc Fe used in the bulk-like 
approximation of the dislocation force constants (see Sec.~\ref{subsec:DFT-FBC}). To ensure that the LGFs computed from the force constants match the 
elastic Green function in the limit ${\mathbf{R}} \rightarrow \infty$, the elastic constants $C_{ijkl}$ computed from the bulk force constant matrix 
$D_{ij}({\mathbf{R}})$ must match the elastic constants computed using standard stress-strain calculations~\cite{trinkle08}. The elastic constants 
$C_{ijkl}$ of a crystal with a single basis atom can be computed from the force constant matrix $D_{ij}({\mathbf{R}})$ using the method of long 
waves~\cite{trinkle08, born54},
\begin{equation}\label{eqn:Cijkl-from-Dij}
-\sum_{\mathbf{R}} D_{ij}({\mathbf{R}}) R_k R_l = V_0 \left( C_{ikjl} + C_{iljk} \right),
\end{equation}
where $V_0$ is the volume of the primitive cell. However, numerical errors in the DFT forces between pairs of atoms with large ${\mathbf{R}}$ can 
compound to produce large errors in the $C_{ijkl}$. We examine the effect of supercell size on the errors in the force constants and the 
corresponding computed $C_{ijkl}$ by performing small displacement method calculations using $3 \times 3 \times 3$, $4 \times 4 \times 4$, $5 \times 
5 \times 5$, and $6 \times 6 \times 6$ supercells with $10 \times 10 \times 10$, $8 \times 8 \times 8$, $6 \times 6 \times 6$, and $6 \times 6 \times 
6$ $\Gamma$-centered Monkhorst-Pack $k$-point meshes\cite{monkhorst76}, respectively. In all these calculations, the atom at the origin of the 
supercell was given a displacement of 0.02 {\AA} along a supercell lattice vector and the resulting forces were input into the code 
{\sc{phon}}~\cite{alfe09} to compute the force constants. We find that the force constants computed using the $4 \times 4 \times 4$ supercell produce 
$C_{ijkl}$ values closest to the $C_{ijkl}$ from stress-strain calculations~\cite{Fellinger2017}, but the values differ by up to 25 GPa. We therefore 
computed the force constants of bulk bcc Fe using the force data from the $4 \times 4 \times 4$ supercell calculation under the constraint that the 
sum in Eqn.~\ref{eqn:Cijkl-from-Dij} gives $C_{ijkl}$ values that exactly match the $C_{ijkl}$ from our stress-strain calculations. These constrained 
force constants are used in the bulk-like approximation of the force constants for the $a_0[100](010)$ edge, $a_0[100](011)$ edge, and 
$a_0/2[111](1\bar{1}0)$ $71^{\circ}$ mixed dislocations. Figure~\ref{fig:FC-vs-vol-V9} compares the unconstrained force constants under volumetric
strain computed with GAP and DFT using $6 \times 6 \times 6$ supercells. Force constants computed using classical potentials like GAP are not subject
to the same types of numerical error as the DFT force constants, so we do not constrain the GAP force constants computed directly for the 
$a_0/2[\bar{1}\bar{1}1](1\bar{1}0)$ edge dislocation (see Sec.~\ref{subsec:DFT-FBC}).

We use DFT with FBC to relax the atoms in regions 1 and 2 of the edge and mixed dislocation geometries. We sample the Brillouin zones of the 
dislocation supercells using $1 \times 1 \times 18$, $1 \times 1 \times 14$, $1 \times 1 \times 8$, and $1 \times 1 \times 20$ $\Gamma$-centered 
Monkhorst-Pack meshes for the $a_0[100](010)$ edge, $a_0[100](011)$ edge, $a_0/2[\bar{1}\bar{1}1](1\bar{1}0)$ edge, and $a_0/2[111](1\bar{1}0)$ 
$71^{\circ}$ mixed dislocations, respectively. We relax the atoms in regions 1 and 2 of the $a_0[100](010)$ edge, $a_0[100](011)$ edge, and 
$a_0/2[111](1\bar{1}0)$ $71^{\circ}$ mixed dislocation geometries until the forces on the ions are less than 5 meV/{\AA}. Due to the larger 
computational cost of relaxing the $a_0/2[\bar{1}\bar{1}1](1\bar{1}0)$ edge dislocation, we relax the atoms in regions 1 and 2 of this dislocation 
until all of the forces on the ions are less than 18 meV/{\AA}. We compared the final relaxed core structures of the other dislocations to their core 
structures earlier in their relaxation when the largest forces were $\sim 18$ meV/{\AA}, and found negligible differences in the geometries; 
therefore, we consider the $a_0/2[\bar{1}\bar{1}1](1\bar{1}0)$ edge dislocation core structure to effectively be fully optimized by that point in the 
relaxation. 

\subsection{Dislocation core visualization}\label{subsec:DD-Nye}

We visualize the relaxed core structures of the dislocations using a combination of differential displacement (DD) maps~\cite{vitek70}, Nye tensor 
components $\alpha_{jk}$~\cite{hartley05a, hartley05b}, volumetric strain $e_V$, and changes in the local magnetic moments $m$ on the Fe atoms. The 
DD maps display the core structure of a dislocation as arrows that indicate the relative displacements between pairs of atoms. The Nye tensor 
components $\alpha_{jk}$ represent the local Burgers vector density at each site in the dislocation core, where the first index $j$ corresponds 
to the dislocation threading direction and the second index $k$ specifies the Cartesian component of the local Burgers vector at each site. For the 
dislocations in this study, the only non-zero Nye tensor components are $\alpha_{3k}$ since the threading direction of each dislocation is chosen 
along the $z$-axis. We visualize the Nye tensor distributions as linearly interpolated contour plots. The dislocations strain the lattice, and 
magnetostrictive materials such as Fe show changes in magnetism under strain~\cite{lacheisserie83}. The dislocation strain fields and the 
corresponding local changes in the magnetic moments on the Fe atoms give a complementary view of the core structures.

We define the centers and widths of the dislocation cores as the first and second moments of the Nye tensor distributions. We define the 
normalized Nye tensor components $\widetilde{\alpha}_{3k}$ as
\begin{equation}\label{eqn:Nye-normalized}
\widetilde{\alpha}_{3k}(x,y) := \frac{|\alpha_{3k}(x,y)|}{\sum_{x',y'} |\alpha_{3k}(x',y')|},
\end{equation}
where $(x,y)$ is the coordinates of a site in the plane normal to the dislocation threading direction. The 
first moments $\overline{x}_{3k}$ and $\overline{y}_{3k}$ of the normalized Nye tensor components,
\begin{equation}
\begin{split}
\overline{x}_{3k} &:= \sum_{x,y} x\, \widetilde{\alpha}_{3k}(x,y),\\
\overline{y}_{3k} &:= \sum_{x,y} y\, \widetilde{\alpha}_{3k}(x,y),
\end{split}
\label{eqn:Nye-1st-moment}
\end{equation}
define the center of each $\alpha_{3k}$ distribution. The second moments $\sigma_{3k,x}$ and 
$\sigma_{3k,y}$ of the normalized Nye tensor components,
\begin{equation}
\begin{split}
\sigma^2_{3k,x} &:= \sum_{x,y} \left(x - \overline{x}_{3k}\right)^2 \widetilde{\alpha}_{3k}(x,y),\\
\sigma^2_{3k,y} &:= \sum_{x,y} \left(y - \overline{y}_{3k}\right)^2 \widetilde{\alpha}_{3k}(x,y),
\end{split}
\label{eqn:Nye-2nd-moment}
\end{equation}
give the widths of a Nye tensor distribution.

We compute Fourier coefficients of the Nye tensor distributions to quantify the symmetry differences between the dislocation core structures 
computed using DFT and the core structures computed using different classical potentials.
The $p^{\mathrm{th}}$ Fourier coefficient 
$c_{3k,p}$ of each $\alpha_{3k}$ about the center $(\overline{x}_{3k}, \overline{y}_{3k})$ is
\begin{equation}\label{eqn:Nye-Fourier}
c_{3k,p} := \sum_{x,y} \alpha_{3k}(x,y) e^{-i p \theta(x,y)},
\end{equation}
where $\theta(x,y):=\arctan\left[(y-\overline{y}_{3k})/(x-\overline{x}_{3k})\right]$ is the angular coordinate of a site $(x,y)$. The $c_{3k,p}$ quantify the $p$-fold rotational symmetry content of the
Nye tensor distributions.

Lastly, we compute the local volumetric strain at each site near the dislocation cores using~\cite{yasi10}
\begin{equation}\label{eqn:local-strain}
e_V := \left[ \frac{\mathrm{det} \big\{\sum_{\mathbf{v'}} v'_j v'_k \big\} }{\mathrm{det} \big\{\sum_{\mathbf{v}} v_j v_k \big\} } \right]^{1/2} - 1,
\end{equation}
where $\mathbf{v'}$ are the nearest neighbor vectors of an atom in the dislocation geometry, $\mathbf{v}$ are the corresponding nearest neighbor vectors
in bulk, and $j$ and $k$ denote Cartesian components. Since the strain is computed at discrete sites like the Nye tensor components, we visualize
the strain distributions as linearly interpolated contour plots.

\section{Results}\label{sec:Results}

\subsection{Dislocation core structures: First-principles calculations}

\begin{figure*}[htb]
\centering
\includegraphics[width=6.00in]{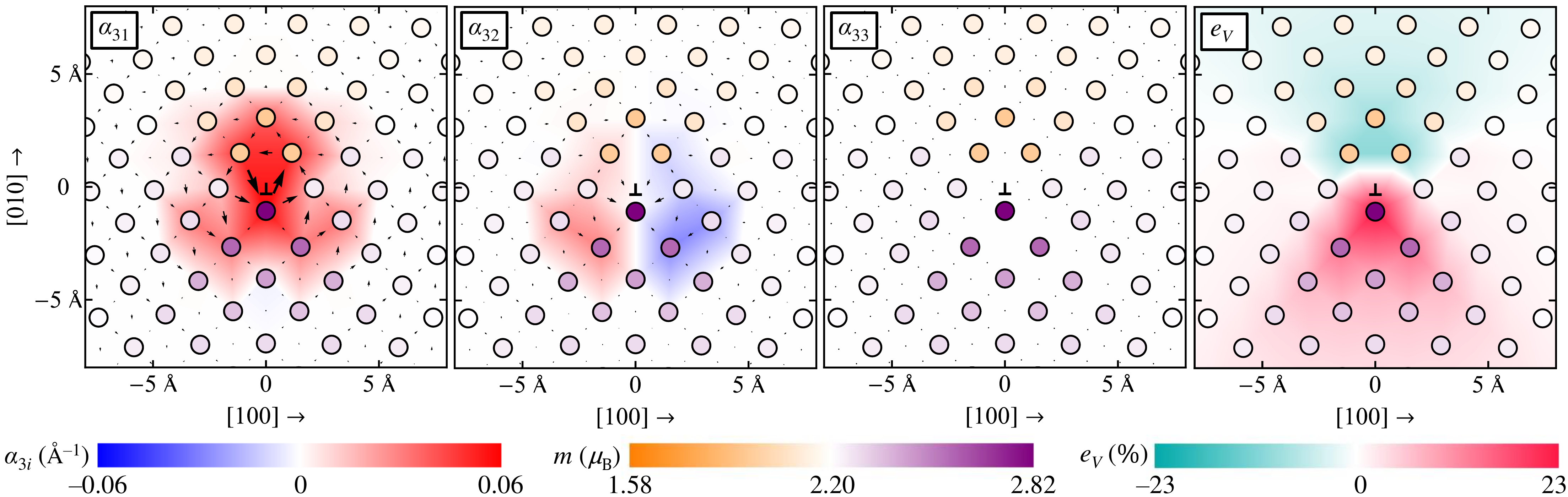}
\caption{(color online). Core structure of the $a_0[100](010)$ edge dislocation in bcc Fe. The first three panels show the differential displacement 
maps using black arrows and the Nye tensor components $\alpha_{3i}$ as contour plots (blue to red color scale). The $\alpha_{31}$ and $\alpha_{32}$ 
distributions reflect the edge character of the dislocation, and the $\alpha_{33}$ distribution reflects the screw character. The 
fourth panel shows the volumetric strain $e_V$ as a contour plot (cyan to magenta color scale). The atoms in all four panels are colored based on 
their magnetic moments $m$ (orange to purple color scale). The core has edge character in both the $x$- and $y$-directions and it remains compact 
after relaxation. The screw component $\alpha_{33}$ of the dislocation is zero. The magnetic moments on the Fe atoms decrease in the compressive 
region above the slip plane and increase in the tensile region below the slip plane.}
\label{fig:edge100}
\end{figure*}

\begin{figure*}[htb]
\centering
\includegraphics[width=6.00in]{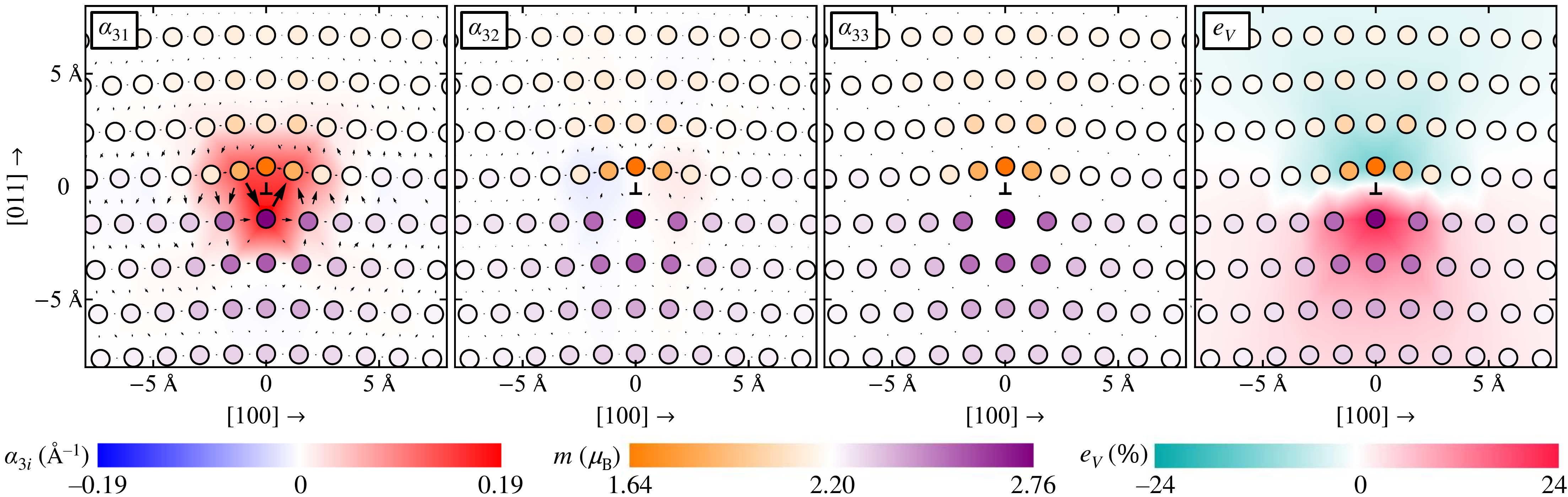}
\caption{(color online). Core structure of the $a_0[100](011)$ edge dislocation in bcc Fe. Similar to the $a_0[100](010)$ edge dislocation, the core
is compact, the screw component is zero, and the magnetic moments decrease(increase) if the atoms are above(below) the slip plane due to the 
dislocation strain field. The edge component of this dislocation in the $y$-direction is nearly zero.}
\label{fig:edge110}
\end{figure*}

\begin{figure*}[htb]
\centering
\includegraphics[width=6.00in]{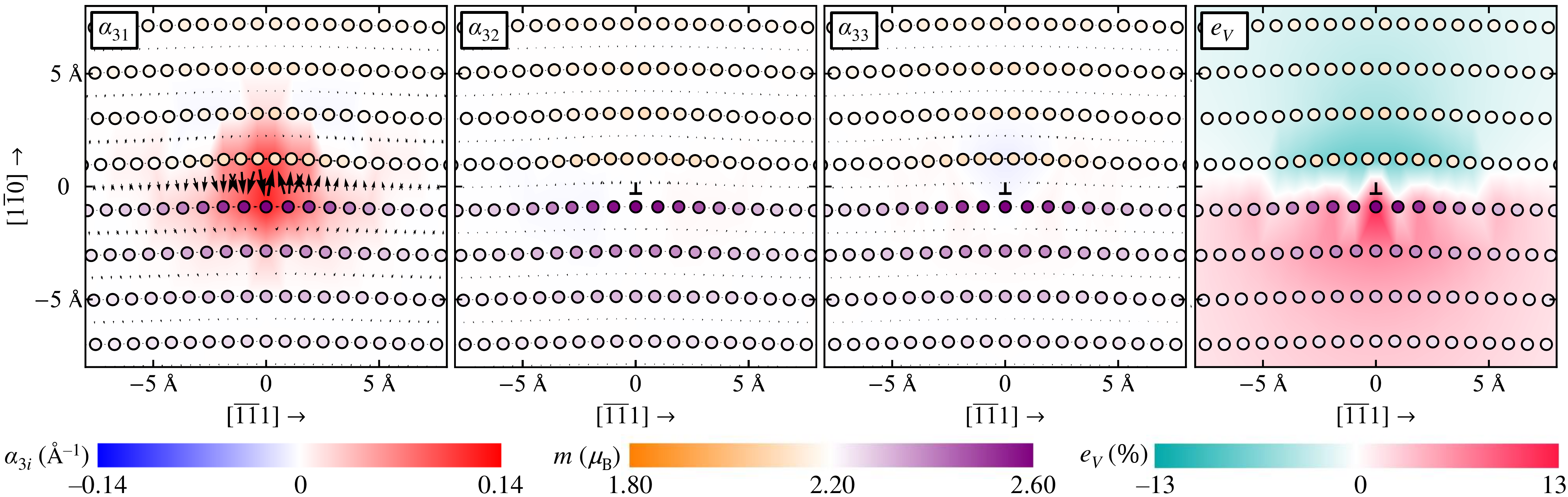}
\caption{(color online). Core structure of the $a_0/2[\bar{1}\bar{1}1](1\bar{1}0)$ edge dislocation in bcc Fe. Similar to the $a_0[100](010)$ and 
$a_0[100](011)$ edge dislocations, the relaxed core is compact, the screw component is zero, and the magnetic moments increase(decrease) in response
to compressive(tensile) strains in the core. The edge component of this dislocation in the $y$-direction is nearly zero.}
\label{fig:edge112}
\end{figure*}

Figures~\ref{fig:edge100}--\ref{fig:edge112} show that the DFT-optimized core structures of the edge dislocations are compact and the magnetic 
moments on the atoms above (below) the slip planes decrease (increase) due to the volumetric strain fields around the dislocation cores. The 
$\alpha_{32}$ and $\alpha_{33}$ distributions are nearly zero for the $a_0[100](011)$ and $a_0/2[\bar{1}\bar{1}1](1\bar{1}0)$ edge dislocations, but 
unexpectedly we find that $\alpha_{32}$ is about one-half as large as $\alpha_{31}$ for the $a[100](010)$ edge dislocation. The $x$- and 
$y$-directions for the $a[100](010)$ dislocation are both $\langle 100 \rangle$-type directions, and we surmise that it is more energetically 
favorable to displace in the $y$-direction compared to the other two edge dislocations. Separately, we have optimized the core structure of the 
$a_0/2[111]$ screw dislocation in bcc Fe using FBC~\cite{fellinger18}. The relaxed core structure is symmetric and compact like in other bcc 
metals~\cite{ismail2000,woodward2001,woodward2002,frederiksen03,Ventelon2013,dezerald2015}, and we compute the widths of the core as 
$2\sigma_{33,x}=3.20$ {\AA} and $2\sigma_{33,y}=3.25$ {\AA}. Table~\ref{tab:Nye-moments} shows that the widths of the edge dislocation cores are 
similar to the widths of the screw dislocation core, confirming that the edge dislocation cores remain compact after relaxation. The $\alpha_{31}$ 
and $\alpha_{32}$ distributions of the $a_0[100](010)$ edge dislocation go to zero at similar distances from their centers, but $\alpha_{32}$ has a 
larger $x$ width since it is antisymmetric. The fourth panels in Figures~\ref{fig:edge100}--\ref{fig:edge112} illustrates the magnetostrictive effect 
in the dislocation cores---compressive strain reduces magnetization and tensile strain increases magnetization. We initialize the magnetic moments 
for all four dislocations in this study in a ferromagnetic state with equal moment values. The relaxed moment values decrease or increase based on 
the local strain distribution, but the ordering remains ferromagnetic throughout all four geometries. We further explore the changes in magnetic 
moments later in this section (see Figure~\ref{fig:momentsVSnndist}).

\begin{table}
\caption{\label{tab:Nye-moments} Widths of the Nye tensor distributions $\alpha_{3k}$ for the edge and mixed
dislocations in bcc Fe. We define the widths of $\alpha_{3k}$ in the $x$- and $y$-directions as two times the corresponding second moment 
computed using Eqn.~\ref{eqn:Nye-2nd-moment}. The edge and mixed cores are compact since their widths are comparable to the widths of 
$\alpha_{33}$ for the ${a_0}/2[111]$ screw  dislocation in bcc Fe ($x$ width = 3.20 {\AA} and $y$ width = 3.25 {\AA}).}
\centering
\begin{tabular}{ccc} \hline\hline
dislocation, $\alpha_{3k}$                                & $x$ width ({\AA}) & $y$ width ({\AA}) \\ \hline
$a_0[100](010)$ edge, $\alpha_{31}$                       & 3.78              & 3.92              \\
$a_0[100](010)$ edge, $\alpha_{32}$                       & 4.69              & 3.04              \\
$a_0[100](011)$ edge, $\alpha_{31}$                       & 4.31              & 3.38              \\
$a_0/2[\bar{1}\bar{1}1](1\bar{1}0)$ edge, $\alpha_{31}$   & 4.33              & 3.00              \\
$a_0/2[111](1\bar{1}0)$ $71^{\circ}$ mixed, $\alpha_{31}$ & 3.97              & 3.28              \\
$a_0/2[111](1\bar{1}0)$ $71^{\circ}$ mixed, $\alpha_{33}$ & 4.41              & 3.30              \\ \hline\hline
\end{tabular}
\end{table}

\begin{figure*}[htb]
\centering
\includegraphics[width=6.00in]{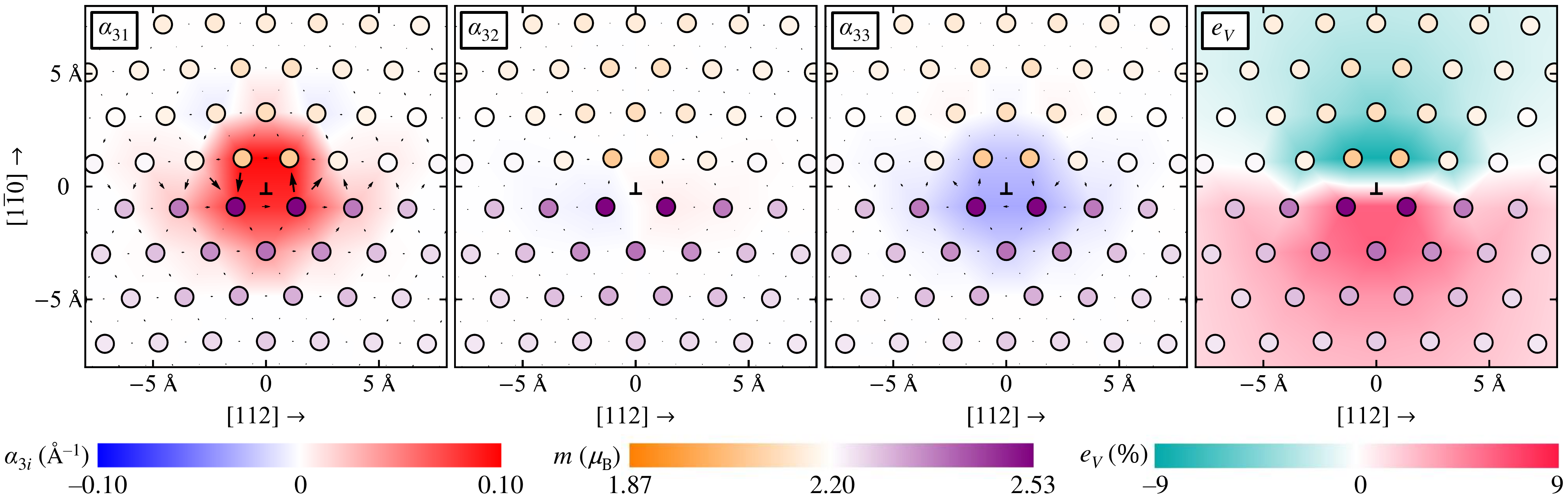}
\caption{(color online). Core structure of the $a_0/2[111](1\bar{1}0)$ $71^{\circ}$ mixed dislocation in bcc Fe. In this
case, the dislocation has both edge ($\alpha_{31}$) and screw ($\alpha_{33}$) components due to its mixed character. Since 
the volumetric strain due to the screw component is small, the changes in the magnetic moments of the Fe atoms are largely 
due to the edge component of the dislocation. The dislocation core is compact after relaxation like the edge dislocation cores.}
\label{fig:mixed}
\end{figure*}

\begin{figure}[htb]
\centering
\includegraphics[width=3.00in]{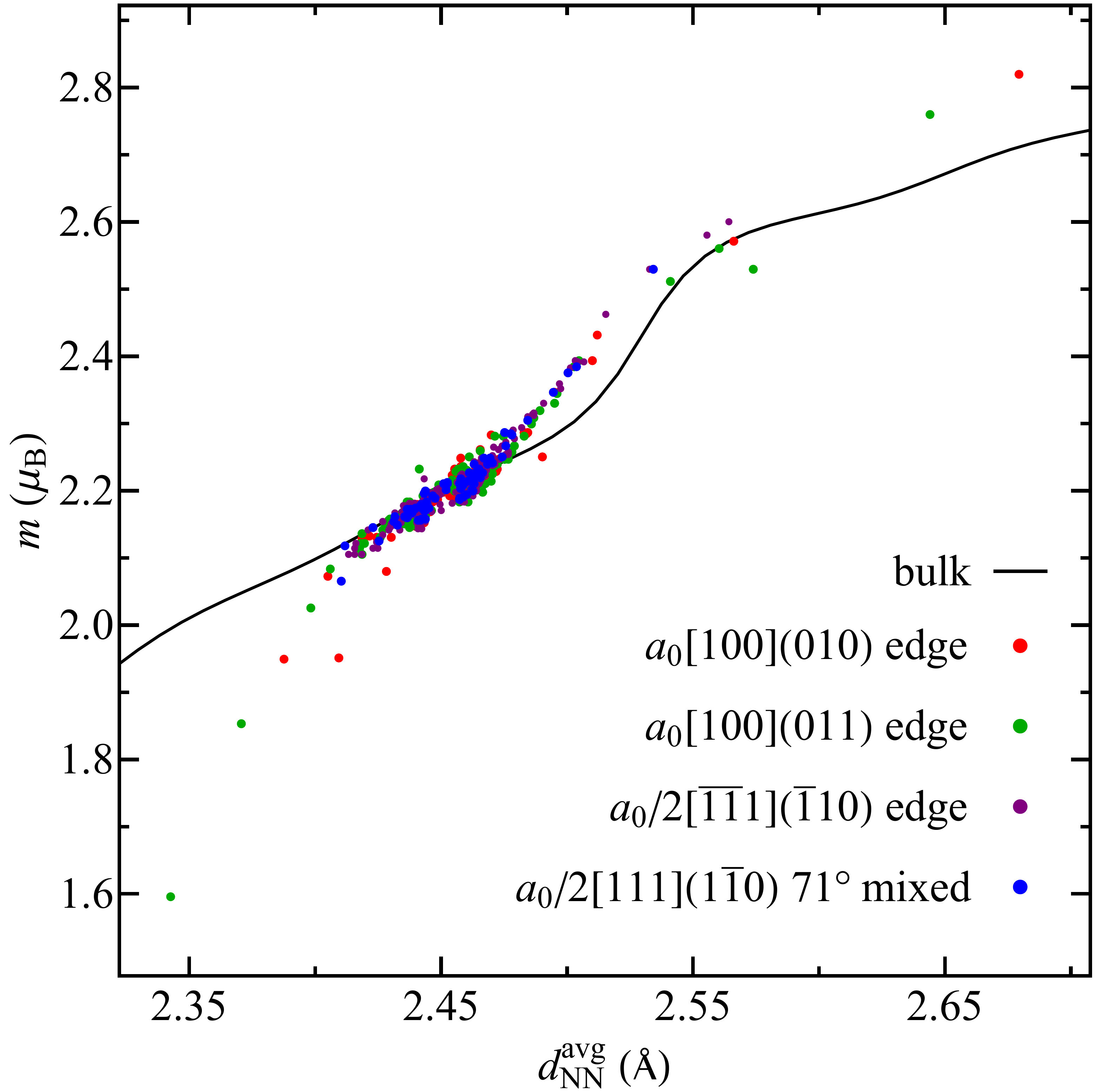}
\caption{(color online). Local magnetic moments $m$ near the dislocation cores versus average nearest-neighbor distance 
$d^{\mathrm{avg}}_{\mathrm{NN}}$. The discrete points are the values for the magnetic moments near the dislocation cores and the solid line show the 
variation of the magnetic moment of bulk bcc Fe versus nearest-neighbor distance. The average nearest-neighbor distance is an alternative measure of 
local volumetric strain which better correlates the magnetic moments near the dislocations with the moments in strained bulk, especially for the 
large strains found in the dislocation cores. The average nearest-neighbor distance in unstrained bulk bcc Fe is 
$d^{\mathrm{avg}}_{\mathrm{NN}} =2.453$ {\AA}.}
\label{fig:momentsVSnndist}
\end{figure}

Figure~\ref{fig:mixed} shows that the DFT-optimized core structure of the $a_0/2[111](1\bar{1}0)$ $71^{\circ}$ mixed dislocation is compact and the 
changes in the magnetic moments on the atoms near the core reflect the volumetric strain field of the edge component. 
The Burgers vector and threading 
direction for the mixed dislocation are along two different body-diagonals of the cubic unit cell, separated by an angle of $\approx 71^{\circ}$. 
Hence, the edge component $\alpha_{31}$ of the dislocation is larger than the screw component $\alpha_{33}$ as shown in Figure~\ref{fig:mixed}. The 
edge component perpendicular to the Burgers vector $(\alpha_{32})$ is nearly zero. Similar to the edge dislocations, the magnetic moments on atoms 
above the slip plane are reduced from their bulk values due to compressive strain and the moments on the atoms below the slip plane are enhanced due 
to tensile strain. This is primarily due to the volumetric strain field generated by the edge component of the dislocation (see 
Fig.~\ref{fig:momentsVSnndist}), since the volumetric strain induced by the screw component is small.

Figure~\ref{fig:momentsVSnndist} shows that the magnetic moments around the dislocation cores closely follow the magnetic moments in bulk bcc Fe for 
small volumetric strains but deviate for the larger strains found in the cores. We use the average nearest-neighbor distance as an alternative 
measure of local volumetric strain since it better correlates the magnetic moments near the dislocations with the moments in strained bulk. For 
reference, the average nearest-neighbor distance in unstrained bulk bcc Fe is $\sqrt{3} a_0/2 = 2.453$ {\AA}. We compute the bulk magnetic moments 
by applying different volumetric strains to the bcc unit cell. However, each dislocation is under a different strain condition since the
normal strain along their different threading directions is zero.
We have also computed the variation in 
magnetization of bulk bcc Fe under the different strain conditions corresponding to each dislocation and found that the behavior is nearly 
identical to the volumetric strain dependence for the strain range shown in the figure. We find that the magnetic moments on the atoms in the 
dislocations closely follow the magnetic moments in strained bulk for sites with about $-2\%$ to $+5\%$ local volumetric strain. The outlying data points 
correspond to atoms right in the dislocation cores where the local strains are larger and non-volumetric contributions to strain may become important.

\subsection{Dislocation core structures: Comparison of interatomic potentials to DFT}

\begin{figure}[htb]
\centering
\includegraphics[width=3.in]{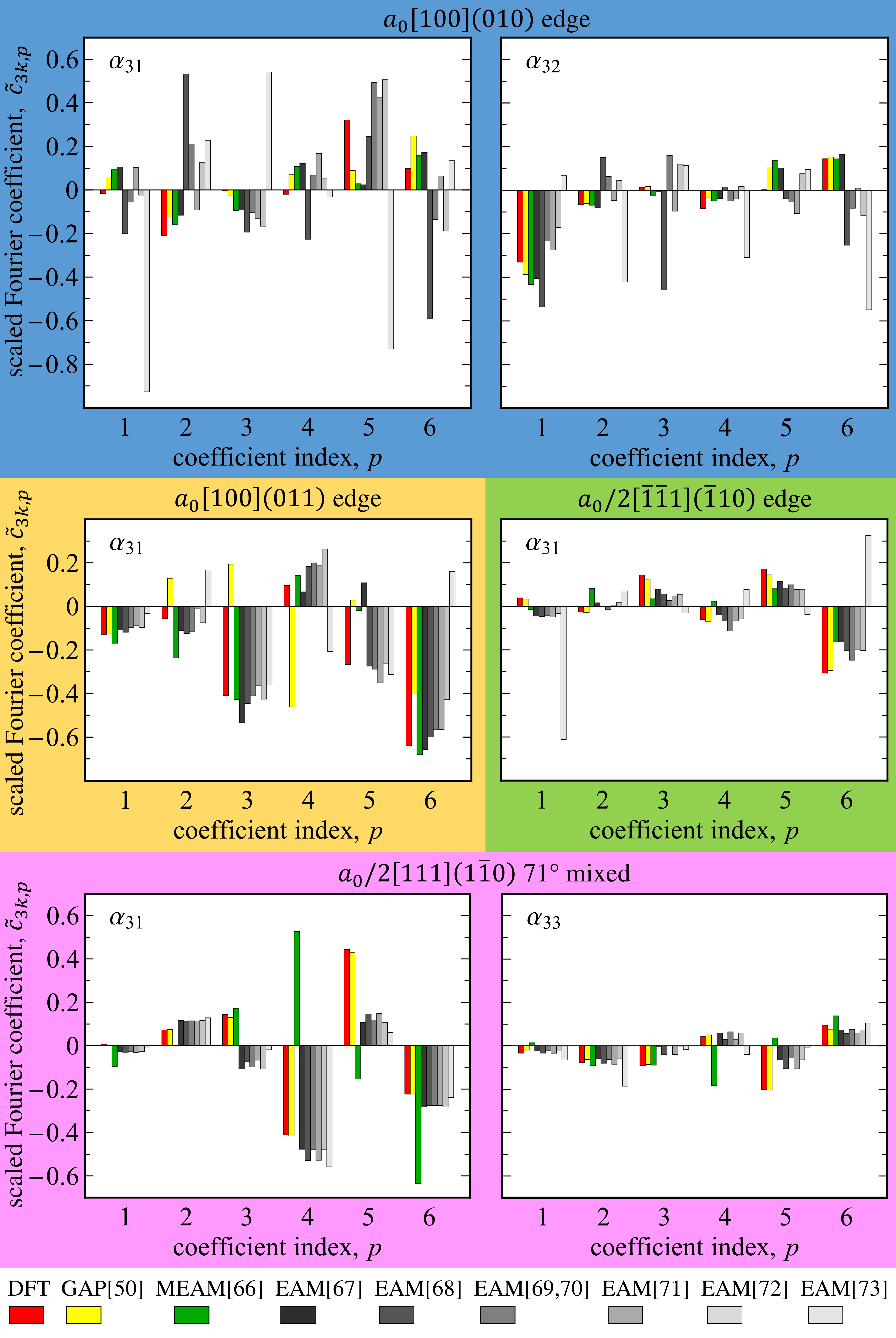}
\caption{(color online). Fourier coefficients of the Nye tensors computed using DFT and GAP~\cite{dragoni18}, MEAM~\cite{asadi15}, and 
EAM~\cite{ackland97, ramasubramaniam09, malerba10, maranica12, mendelev03, chamati06, proville12} potentials. The coefficients with even indices are 
real and the coefficients with odd indices are imaginary. We show only the coefficient values for positive indices since the negative even 
coefficients equal the positive even coefficients, and the negative odd coefficients have equal magnitudes and opposite signs as the positive odd 
coefficients. The scaled coefficients in the figure are defined as ${\widetilde{c}}_{3k,p} = c_{3k,p}/c_{31,0}$. The plots reveal the differences in 
symmetry between the cores, and can be used to quickly judge if a given potential produces a core structure similar to DFT. For example, the GAP 
cores of the $a_0[100](010)$ edge, $a_0/2[\bar{1}\bar{1}1](1\bar{1}0)$ edge, and $a_0/2[111](1\bar{1}0)$ $71^{\circ}$ mixed dislocation agree well with 
DFT, but the GAP core of the $a_0[100](011)$ edge dislocation relaxes to a more open structure (see Fig.~\ref{fig:t011-compare-a31} for a direct 
comparison of the cores).}
\label{fig:Nye-Fourier}
\end{figure}

\begin{figure*}[htb]
\centering
\includegraphics[width=6.in]{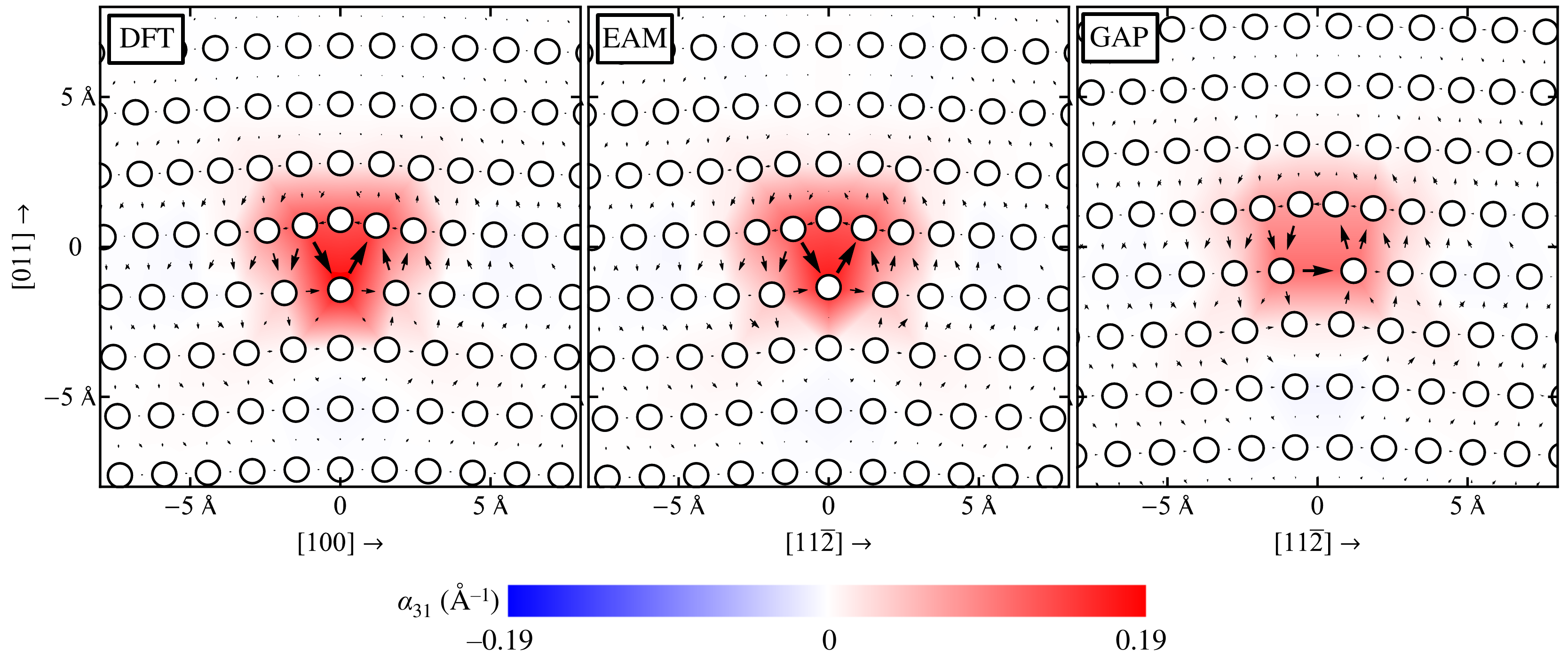}
\caption{(color online). Comparison of DFT, EAM~\cite{mendelev03}, and GAP~\cite{dragoni18} results for the core structure 
of the $a_0[100](011)$ edge dislocation. The figures show the $\alpha_{31}$ edge component of the Nye tensor, and the atoms in this DFT figure are not 
colored based on their magnetic moments. The EAM potential produces a core structure similar to DFT, but the GAP core is different. This is reflected
by the large differences between the DFT and GAP Fourier coefficients in Fig.~\ref{fig:Nye-Fourier}}.
\label{fig:t011-compare-a31}
\end{figure*}

Figure~\ref{fig:Nye-Fourier} compares the DFT core structures of the edge and mixed dislocations to the cores from GAP~\cite{dragoni18}, 
MEAM~\cite{asadi15}, and EAM~\cite{ackland97, ramasubramaniam09, malerba10, maranica12, mendelev03, chamati06, proville12} potentials using the 
Fourier coefficients $c_{3k,p}$ of the Nye tensor distributions. The classical potential calculations are performed using the code 
{\sc{lammps}}~\cite{plimpton95}, with potential parameters downloaded from the NIST Interatomic Potential Repository~\cite{NISTpotentials} with the 
exception of Ref.~\cite{ramasubramaniam09} EAM, which used the recommended \texttt{PotentialB.fs} file downloaded from Ref.~\cite{carterSITE}. The 
supercells in the classical potential calculations contain cylindrical slab geometries with approximately 20,000 atoms surrounded by vacuum. We use 
fixed boundary conditions where the atoms at a distance less than the potential cutoff radii from the vacuum are held at their positions from 
anisotropic elasticity theory while all the other atoms are relaxed using a conjugate gradient method. The $c_{3k,p}$ (see 
Eqn.~\ref{eqn:Nye-Fourier}) quantify the differences in the $p$-fold symmetry content between the dislocation cores computed using different methods. 
For example, the core of the $a_0[100](011)$ edge dislocation relaxes to a different structure than the DFT core using the GAP and there are large 
difference between the GAP and DFT $c_{31,p}$ for $p > 1$. In contrast, the EAM and MEAM $c_{31,p}$ for this dislocation agree well with the DFT 
values. Figure~\ref{fig:t011-compare-a31} shows that the core computed using the EAM potential from Ref.~\cite{mendelev03} is similar to the DFT 
core, but the GAP core relaxes to a more open structure. We find the largest differences from the DFT core structures when the $a_0[100](010)$ edge 
dislocation is relaxed using the EAM potentials from Refs.~\cite{ramasubramaniam09, proville12}, when the $a_0[100](011)$ edge 
dislocation is relaxed using GAP~\cite{dragoni18}, when the $a_0/2[\bar{1}\bar{1}1](1\bar{1}0)$ edge dislocation is relaxed using the EAM potential 
from Ref.~\cite{proville12}, and when the $a_0/2[111](1\bar{1}0)$ $71^{\circ}$ mixed dislocation is relaxed using the MEAM potential~\cite{asadi15}. 
The study in Ref.~\cite{haghighat14a} found that the EAM potential from Ref.~\cite{mendelev03} produces a different core structure for $a_0/2\langle 
111 \rangle \{110\}$ edge dislocations compared to the EAM potentials in Refs.~\cite{ackland97,malerba10,maranica12}, whereas we find that all of 
these potentials produce core structures similar to our DFT core. We are able to reproduce the core structures in Ref.~\cite{haghighat14a} by 
choosing different elastic centers for the initial dislocation geometry, but these cores transform to the other core after annealing from 300K. We 
also find that the two types of cores are nearly degenerate in energy which is consistent with the nudged elastic band calculations in 
Ref.~\cite{haghighat14a}, so it is likely that the core we found is the ground state structure and the other core is a transition state as the 
dislocation moves in its slip plane.

The alternate structure of the $a_0/2[\bar{1}\bar{1}1](1\bar{1}0)$ edge dislocation for the EAM potential from Ref.~\cite{mendelev03} discussed in 
the last paragraph raises the question about the existence of metastable states for the other dislocation cores considered in this study. Metastable 
core structures are most likely for dislocations with large spreading in the slip plane or that dissociate into partial dislocations separated by 
stacking fault since multiple energy minimia are present in the slip plane. We do not expect metastable core structures to exist for the dislocations 
in this study since all the DFT cores are compact. We invesitagate this idea further by annealing the cores from the EAM and MEAM potentials that are 
most similar to the DFT cores to examine if these structures are stable. We anneal the $a_0[100](010)$ edge dislocation cores for the EAM potentials 
from Refs.~\cite{ackland97, mendelev03} and the MEAM potential, the $a_0[100](011)$ edge cores for the EAM potentials from 
Refs.~\cite{ackland97,ramasubramaniam09,malerba10,maranica12,mendelev03,chamati06} and the MEAM potential, the $a_0/2[\bar{1}\bar{1}1](1\bar{1}0)$ 
edge cores for the EAM potentials from Refs.~\cite{ackland97,ramasubramaniam09,malerba10,maranica12,mendelev03,chamati06} and the MEAM potential, and 
the mixed cores for the EAM potentials from Refs.~\cite{ackland97,ramasubramaniam09,malerba10,maranica12,mendelev03,chamati06}. In each case, the 
initial geometry for the annealing simulation is the conjugate gradient-optimized geometry with Fourier coefficients shown in Figure 8. We anneal the 
cores from a starting temperature of 300K and then perform a subsequent conjugate gradient geometry optimization. All of the annealed core structures 
remain unchanged except for the $a_0[100](010)$ edge dislocation from the EAM potential in Ref.~\cite{mendelev03} which remains compact but becomes 
asymmetric in the slip direction, the $a_0[100](011)$ edge dislocation from the MEAM potential which has a larger spreading in the slip plane than 
the initial structure, and the mixed dislocation from the EAM potential in Ref.~\cite{ackland97} which transforms to a structure similar to the MEAM 
structure. The GAP cores of the $a_0[100](010)$ edge, $a_0/2[\bar{1}\bar{1}1](1\bar{1}0)$ edge, and $a_0/2[111](1\bar{1}0)$ $71^{\circ}$ mixed 
dislocations are similar to the DFT cores. GAP calculations are more computationally expensive than EAM and MEAM calculations, so we only annealed 
the GAP mixed dislocation core. For the two GAP edge dislocations that are similar to DFT we applied small random displacements to the atoms in the 
core region and then relaxed the geometry using a conjugate gradient method. All three GAP dislocation cores relax back to their starting geometries. 
Finally, we investigated the stability of the DFT mixed dislocation geometry by performing restoring force calculations. We added small displacements 
along the slip direction to the four atoms directly above the slip plane that are closest to the center of the dislocation core, and computed the 
resulting forces using DFT. The forces primarily point opposite to the displacement direction, indicating that the core will relax back to the 
original geometry. All of these test calculations strongly suggest that the DFT core structures reported in this study are stable groundstate 
structures, and that the core transformations we find after annealing are due to artifacts in the interatomic potentials. None of the potentials is 
able to produce core geometries similar to DFT for all of the dislocations, but the EAM potential from Ref.~\cite{mendelev03} has the best overall 
performance. All of the core geometries optimized with this potential using a conjugate gradient method are similar to DFT, and they all remain 
stable under annealing except for the $a_0[100](010)$ edge dislocation which breaks symmetry but remains compact. This EAM potential also produces a 
compact and symmetric core structure for $a_0/2 \langle 111 \rangle$ screw dislocations similar to DFT~\cite{malerba10}.

\section{Summary and discussion}\label{sec:Summary}

We use density functional theory (DFT) with lattice flexible boundary conditions (FBC) to optimize the core structures of $a_0[100](010)$ edge, 
$a_0[100](011)$ edge, $a_0/2[\bar{1}\bar{1}1](1\bar{1}0)$ edge, and $a_0/2[111](1\bar{1}0)$ $71^{\circ}$ mixed dislocations in bcc Fe. The FBC 
approach couples the highly-distorted dislocation core which is treated with DFT to an infinite harmonic lattice via the lattice Green function 
(LGF), which allows the dislocation to effectively relax as an isolated defect. In contrast to most previous first-principles FBC calculations of 
dislocation cores that use the bulk LGF to relax the harmonic region outside the core, we use LGFs specifically computed for each dislocation 
geometry. The simple bulk-like approximation we used for generating the force constants and corresponding LGFs for the $a_0[100](010)$ edge, 
$a_0[100](011)$ edge, and $a_0/2[111](1\bar{1}0)$ $71^{\circ}$ mixed dislocations fails to produce an adequate LGF for the 
$a_0/2[\bar{1}\bar{1}1](1\bar{1}0)$ edge dislocation. For this case, we found that a Gaussian approximation potential (GAP) for bcc Fe produces 
accurate force constants under strain which lead to a dislocation LGF capable of optimizing the core geometry. We find that the cores of all the 
dislocations in this study are compact and the magnetic moments on the atoms in the cores increase in the tensile region below the slip planes and 
decrease in the compressive region above the slip planes. Except for highly distorted sites nearest to the cores, the strain response of the magnetic 
moments on the atoms in the dislocated geometries closely follows the volumetric-strain response of the magnetic moment in bulk bcc Fe. We find that 
the initial ferromagnetic ordering we impose on the magnetic moments in each geometry remains after relaxation, showing that ferromagnetic ordering 
in the cores is at least metastable. Future studies could investigate the impact of different initial magnetic configurations in the dislocation 
cores on their relaxed magnetic states and geometries. We find that most of the core structures computed using the GAP, MEAM, and EAM interatomic 
potentials compare well with the DFT core structures, with a few notable exceptions where the cores relax to different structures. While none of the 
potentials is able to produce core geometries similar to DFT for all of the dislocations, the EAM potential from Ref.\cite{mendelev03} has the best 
overall performance. All of the core geometries optimized with this potential using a conjugate gradient method are similar to DFT, and they all 
remain stable under annealing except for the $a_0[100](010)$ edge dislocation which remains compact but becomes asymmetric along the slip direction. 
Additionally, this EAM potential produces a compact and symmetric core structure for $a_0/2 \langle 111 \rangle$ screw dislocations similar to 
DFT\cite{malerba10}. Relaxed dislocation core structures are of fundamental importance for understanding plasticity in bcc Fe, provide the geometries 
required for first principles-based studies of solid-solution strengthening\cite{yasi10} and solute diffusion near dislocations\cite{schiavone16}, 
provide data for parameterizing and benchmarking more computationally efficient models such as classical interatomic potentials, and serve as a 
comparison point for future experimental measurement of edge and mixed dislocation core structures in bcc Fe.

\section{Data availability}\label{sec:Data}

The {\sc{vasp}} and {\sc{lammps}} input files used to perform the calculations along with the relaxed dislocation core geometries are available to 
download from http://hdl.handle.net/11256/978.

\begin{acknowledgments}
This material is based upon work supported by the Department of Energy National Energy Technology Laboratory under Award Number DE-EE0005976. 
Additional support for this work was provided by NSF/DMR Grant No. 1410596. This report was prepared as an account of work sponsored by an agency of 
the United States Government. Neither the United States Government nor any agency thereof, nor any of their employees, makes any warranty, express or 
implied, or assumes any legal liability or responsibility for the accuracy, completeness, or usefulness of any information, apparatus, product, or 
process disclosed, or represents that its use would not infringe privately owned rights. Reference herein to any specific commercial product, 
process, or service by trade name, trademark, manufacturer, or otherwise does not necessarily constitute or imply its endorsement, recommendation, or 
favoring by the United States Government or any agency thereof. The views and opinions of authors expressed herein do not necessarily state or 
reflect those of the United States Government or any agency thereof. The research was performed using computational resources provided by the 
National Energy Research Scientific Computing Center. Additional computational resources were sponsored by the Department of Energy's Office of 
Energy Efficiency and Renewable Energy and located at the National Renewable Energy Laboratory, the General Motors High Performance Computing Center, 
and the Golub cluster maintained and operated by the Computational Science and Engineering Program at the University of Illinois.
\end{acknowledgments}

\end{document}